# Quantitative analysis of collagen remodeling in pancreatic lesions using computationally translated collagen images derived from brightfield microscopy images

Varun Nair [a], Gavish Uppal [a], Saurav Bharadwaj [a], Ruchi Sinha [b], Manjit Kaur [c], Rajesh Kumar [a]*

[a] Department of Biomedical Engineering, Indian Institute of Technology Ropar, Punjab-140001, India.

[b] Department of Pathology, All India Institute of Medical Sciences Patna, Bihar -801507, India.

[c] Department of Pathology, All India Institute of Medical Sciences Bathinda, Punjab-151001, India.

* Email: rajeshkumar@iitrpr.ac.in

## Abstract

The changes in stromal collagen play a crucial role during the pathogenesis and progression of pancreatic intraepithelial neoplasm (PanIN) to pancreatic ductal adenocarcinoma (PDAC) while misdiagnosis of PanIN is common because of the resemblance to chronic pancreatitis (CP) in its symptoms and subsequent radiographic and histopathological evaluations. To visualize fibrillar collagen in tissues, second harmonic generation (SHG) microscopy is now utilized as a gold standard in various stromal-based research analyses. However, a technical approach that can perform a quantitative analysis of fibrillar collagen directly on standard slides stained with Hematoxylin & Eosin (H&E) can (i) discard the need for specialized and costly equipment or labels, (ii) further supplement the traditional histopathological insights and, (iii) potentially be integrated within the framework of standard histopathology workflow. In this study, the whole-core brightfield H&E-stained images of pancreatic tissues were translated into the new collagen images using a pre-trained convolutional neural network (CNN) based cross-modality image translation algorithm. Subsequently, collagen characteristics of PDAC, PanIN, CP, and normal pancreatic tissues (control) were extracted and compared. The highest alignment ($p < 0.01$, $R^2 = 0.2594$) was observed in PDAC cores in comparison to the remaining three groups, while the lowest fiber density ($p < 0.0001$, $R^2 = 0.3569$) was observed in the case of normal tissue cores. Moreover, the collagen area and fiber length had shown higher area under curve (0.83 and 0.81, respectively) in discriminating neoplastic and non-neoplastic tissues based on their receiver operating characteristics. The study demonstrated that the computationally generated collagen images can provide a quantitative assessment of collagen remodeling in pancreatic lesions. The cross-modality image synthesis using CNN was directly implemented to standard H&E-stained whole core images to evaluate the collagen alterations, which may further lead towards better histopathological and tissue microenvironment insights without the need of specialized imaging equipment or labels.

# Introduction

The tissue microenvironment comprises various cellular components and the extracellular matrix (ECM)[1,2]. The ECM is a dynamic and intricate structure[3] whose chemical and biophysical characteristics influence cell morphology[2], adhesion[4], migration[5], and proliferation[6]. It also governs tissue perfusion[7] and tissue morphogenesis[8]. The majority of the ECM is made up of fibrillar collagen[9], which is rapidly being identified as a key factor in cancer genesis and development[10]. Nearly all major types of cancer have been linked to collagen fiber remodeling[11]. Collagen remodeling is a dynamic process that occurs in response to physiological and pathological stimuli, including tissue damage, inflammation, and fibrosis. Collagen remodeling promotes the formation and progression of tumor microenvironment (TME)[12]. The major phases of collagen remodeling include breakdown, production and crosslinking of fibers, fiber orientation shift, and interactions between cells and stroma[13]. Several studies have demonstrated that the remodeling of collagen could be a significant visual biomarker in numerous clinical pathologies, including malignancies of the breast[14], intestines[15], ovaries[16], prostate[17], and the skin[18]. The onset and development of pancreatic lesions can also be significantly visualized by collagen remodeling[19]. Therefore, it is crucial to comprehend the structural and organizational characteristics of collagen in order to augment clinical diagnostics, predict patient survival, and tailor adjuvant therapy[20].

Various characterization tools, mainly based on intensity derivatives[21] and variation[22], Fourier transform[23], directional filters[24], and fiber tracking algorithms[25], have been developed to study collagen remodeling. Studies have shown the better performance of the curvelet-transform[26] and FIRE algorithm-based tools[27] such as CurveAlign[28] and CT-FIRE[29]. These image analysis tools have been applied to characterize collagen in tandem with various microscopic modalities such as polarized light microscopy (PLM)[30], scanning electron microscopy (SEM)[31], fluorescence microscopy[32], and second harmonic generation (SHG) microscopy[33]. Among these modalities, SHG microscopy[34] provides a label-free and high-resolution visualization of collagen fiber, which is now considered as the gold standard in stromal research. SHG imaging is a second-order nonlinear optical contrast process that enables visualizing biological components such as collagen, myosin, and microtubules at the cellular level in a label-free and non-destructive manner[35,36]. The non-centrosymmetric structure possessed by fibrillar collagen is essential for generating discernible SHG signals[37].

SHG imaging has been used to visualize alterations in collagen morphology and orientation in multiple conditions, such as atherosclerosis[38], osteoarthritis[39], aging[40], wound healing[41], and fibrosis, including pancreatitis, pancreatic intraepithelial neoplasia, and ductal adenocarcinoma[42,43]. In a study by Drifka et.al[44], quantitative SHG microscopy was used to investigate the changes in collagen topology of the periductal stroma of pancreatic ductal adenocarcinoma (PDAC) in comparison with normal pancreatic tissue and chronic pancreatitis (CP) and, was observed that collagen surrounding carcinomic ducts showed increased alignment, length, and width compared to non-neoplastic ducts. Recently, Li et. al[45] demonstrated the use of co-registered SHG and H&E images to differentiate between PDAC and CP. They showed that incorporating the collagen characteristics extracted from SHG images with H&E data increased the accuracy and area under the curve of the receiver operating characteristics of differentiation algorithm. Another clinically relevant SHG-based quantitative study has shown a correlation between high stromal collagen alignment with poor patient prognosis in PDAC[46]. Although SHG microscopy is utilized for quantitative collagen remodeling studies at various levels including the stromal characterization, it is hardly in use under clinical settings owing to its substantial cost, intricacy, and the usual necessity of photonics specialists for operating the device.

Owing to the advent of cross-modality image translation, the disadvantages related to the acquisition of SHG images have been attempted to overcome in the past decade. Image-to-image translation is a technique where deep learning algorithms are employed to convert an image from one domain to another, with the objective of generating an output image that closely resembles the target domain while retaining the important features of the source image[47]. Deep learning utilizes convolutional neural networks (CNNs) to address intricate problems in the realm of image processing. CNNs have been used to translate low-resolution images to high-resolution images[48] (single-image super-resolution) or to translate images from one modality to another. For example, it was used to translate bright field (BF) microscopic images of H&E-stained tissues to virtual immunohistochemistry (IHC) based cytokeratin-stained images[49]. Image-to-image translation can improve the accuracy and efficiency of tasks like segmentation[50], feature extraction[51], and classification[52], leading to the advent of digital pathology. The term 'digital pathology' refers to sophisticated slide-scanning methods, together with algorithms based on artificial intelligence for detecting, segmenting, scoring, and diagnosing digitalized whole-slide images[53] to conduct a clinically relevant comprehensive analysis. As it is known that SHG microscopy is capable to image fibrillar collagen with high specificity[54], an effort was made by Keikhosravi et. al., in which SHG equivalent images were generated via image-to-image translation using the BF images of standard H&E-stained tissues for better visualization of fibrillar collagen[55]. Such an approach may potentially circumvent the disadvantages associated with SHG imaging while enhancing the information on collagen remodeling.

In this study, (1) we computationally translated the whole core BF images of H&E-stained TMA to collagen images and showed that translated image is equivalent to the SHG image. (2) Further, we extracted and quantified the collagen features from the translated images and observed a significant difference in collagen morphology among four classes (normal tissue, CP, PanIN, and PDAC) of pancreatic tissues. (3) Moreover, we compared and analyzed the performance of each extracted collagen feature in differentiating the neoplastic stromal microenvironment. Our study has demonstrated the application of computationally translated whole core images in the quantitative analysis of pancreatic stromal collagen, complementing conventional histopathological insight and paving the way to further explore the potential use of fibrillar collagen as a tissue biomarker in pancreatic lesions.

## Materials and Methods

### a. Tissue microarrays images

The study utilized H&E-stained BF images of commercially available TMA slides originating from US Biomax, Inc. The images were accessed from the available WSISR-TMA image dataset[56] with prior approval. The image dataset was composed of a total of 60 cores from 36 cases. This included 8 cases of PanIN, 8 cases of CP, 4 cases of PDAC, and 4 cases having normal pancreatic tissue (control), in which two cores were acquired from each case. Further, there were 7 cases of CP and 5 cases of PDAC that had single cores per case. The dataset has a balanced gender distribution with 30 cores each representing male and female cases. The median age of the population is 53 years (range: 36 - 76). The TMA slides were sectioned and stored at 4°C before BF images were captured with an Aperio CS2 Digital Pathology Scanner at a 20x magnification (0.504 μm/pixel). The SHG images of H&E-stained TMA cores used for the CNN validation were sourced from a GitHub repository, whose weblink is mentioned in a previously published study[57]. The set comprised 36 SHG images of pancreatic stroma TMA cores that were not included in the test dataset. The diagnostic and demographic information of the tissue core images used for validation were not considered as their associated data is not of primary significance for the same. The SHG images were captured in backward SHG mode at 20x magnification with a custom-built imaging system consisting of an 890nm mode-locked MIRA 900 Ti: Sapphire laser and a 445 ± 20 nm narrow-band pass filter.

### b. Image Preprocessing and Normalization

Image preprocessing needs to be carried out depending on the inherent characteristics of the image to be assessed. The SHG dataset images used for validation of the neural network were rescaled to match the resolution and dimensions of the input BF images. Prior to the comparison with translated output images, we used a two-step RGB intensity-based registration technique to align the H&E-stained images with the SHG images. The red staining of eosin on the stroma was extracted using K-means clustering[58] to produce a false ECM image. An affine transform was found for projecting the false ECM image to the SHG image using a one-plus-one evolutionary algorithm[59]. It is crucial to match the fundamental attributes of the input data with the dataset the neural network was trained on to ensure proper neural network performance. The H&E images of the US Biomax Inc. TMA cores used to examine collagen remodeling in pancreatic lesions displayed variation in stain vectors in comparison to the training dataset. These stain vectors were normalized by histogram modification using the Color Balance function in ImageJ[60].

### c. Neural Network Implementation

The convolutional neural network originally designed and trained by Li et al., was utilized with its pre-existing weights for analysis of pancreatic tissue[55]. The architecture of the network is shown in Fig. 1A. The network was pretrained on 1388 pancreatic TMA H&E-stained BF and SHG image pairs, along with 489 H&E-stained breast biopsy whole slide image pairs. The corresponding weights were loaded into the network for its smooth functionality. The neural network computationally generated the new collagen images from the normalized BF images. The input image was windowed into a size of 128x128px with a 10% overlap to match the size of the input layer of the neural network. A residual module-based encoder-decoder network was employed for the study. For image classification and retrieval, residual mapping provides an effective shallow representation. Using two distinct filter sizes and skip connections enhanced edge preservation for the translation to the collagen image.

To calculate the innate feature space, an input patch is fed into a series of residual modules in the encoder (Fig. 1B). Each residual block comprises two sets of convolutional layers with the exact same number of kernels, each followed by a batch normalization layer and leaky ReLU layer. The two convolutional layers have different filter sizes and strides to decrease the dimensions of the feature maps. This channel's output was combined with the output processed by a residual layer having identical kernel value. The network's inclusion of skip connections (Fig. 1C) significantly increased the training accuracy and efficiency. The architecture of these connections was similar to the encoder

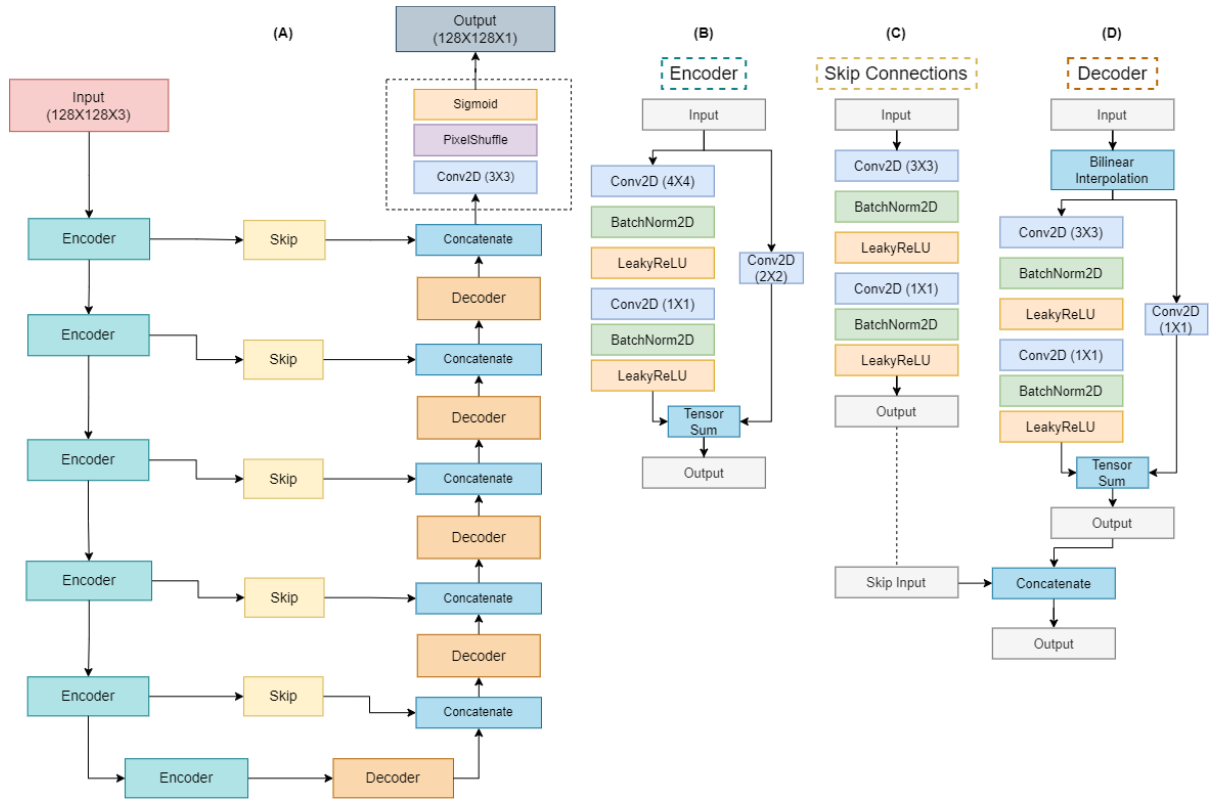

*Figure 1. The network architecture of the implemented neural network (A) with the internal architecture of the encoder block (B), the skip connection (C) and the decoder block (D) [adapted from Keikhosravi et al.,[55]]*

without the inclusion of residual connections. By utilizing bicubic interpolation at every step, coupled with skip connections, the decoder (Fig. 1D) is able to restore the feature space to its original dimensions, matching those of the encoder, while employing a similar residual block structure. Sub-pixel convolution, a combination of group convolution and periodic shuffling, was performed using a pixel shuffle layer[61] for reconstructing the output patch from the feature space of decoder's final layer to a size equivalent to the input patch. The output patches are then stitched using the Stitching plugin[62] in Fiji[63].

To validate the image-to-image translation performance, Structural Similarity Index Measure (SSIM)[64] was computed between each output of the neural network and its corresponding real SHG image (used as the reference image). The 'SSIM Index' plugin was used to capture cognitive measures of the image quality.

### d.  Collagen Fiber Extraction and Quantitative Analysis

The quantitative features of collagen fibers having high specificity and resolution were determined to characterize the alterations in fibrillar collagen of various pancreatic lesions. These features include the length, width, alignment, density, the overall area of collagen deposition, and the dispersion parameter. The CT-FIRE v3.0 tool was utilized to accurately segment and measure the individual collagen fibers. The background noise was removed by thresholding (grayscale 10-255) the computationally generated images and converting them into 8-bit images. After loading the images to CT-FIRE, the fiber extraction parameters were chosen as thresh_im2 (= 20) and s_xlinkbox (= 20), which were set empirically based on visual verification of the detected fibers. The collagen fiber width and length values were quantified in pixels. The alignment and density of fibers were computed using the CurveAlign v5.0 tool. The mean nearest alignment was calculated among the nearest 16 fibers. The alignment of fibers ranges on the scale from 0 to 1, in which 1 denotes the perfect relative alignment. Density was measured as the cumulative mean of the number of fibers inside a moving window of dimensions 128 x 128 pixels.

The area of collagen deposition was defined as the ratio of the collagen pixels in a computationally generated image over whole core pixels in the respective H&E-stained image. To calculate the value, the BF H&E-stained image was first converted into grayscale. Otsu's thresholding[65] was performed on both images to remove background noise. The collagen area (A) was calculated using the formula:

$$A = A_c/A_t,$$

where $A_c$ is the area of the collagen in the translated image and $A_t$ is the area of the whole core tissue. The dispersion parameter 'k' was determined using the tool FiberFit[66] v2.0, where a higher value of 'k' represents the aligned network of fiber. Before applying FiberFit, the computationally generated images were denoised, normalized, and transformed into 8-bit images.

**Statistical Analysis**

GraphPad Prism v9.4.1 was used to conduct the statistical analysis. Bland-Altman diagram[67] was plotted to observe the agreement between the computationally generated collagen images and the SHG images by comparing the difference between measurements of a specific parameter vs. the average of these values. One-way analysis of variance (ANOVA) followed by a pairwise analysis using Fisher's least significant difference (LSD) test was used to compare the collagen parameters. The results were plotted using the bar graphs presented as mean ± standard deviation. The p-values < 0.05 was considered statistically significant. The k-nearest neighbors (KNN)[68] classifier was implemented using scikit-learn library[69] to discriminate between the neoplastic and non-neoplastic tissues. The number of neighbors was iterated from 1 to 20, and stratified 10-fold cross-validation was used to calculate the mean accuracy for each iteration. The final algorithm was trained using the number of neighbors that performed with the highest accuracy. The receiver operating characteristics (ROC) curve was plotted for each parameter based on the prediction probability of the algorithm. The area under the ROC curve was computed to measure the classification performance.

## Results

### a. Performance of Cross-Modality Image Translation

Figure 2 represents the impact of stain normalization on the performance of translated collagen images. The translated output image was not able to accurately represent the distinct collagen fibers observed in the ECM of pre-normalized H&E-stained input image (Fig. 2A-C). To resolve the issue, H&E-stained tissue microarray images were subjected to identical histogram modifications to adjust the stain vectors. The translated collagen image obtained post stain normalization process have shown a significant improvement in detection of collagen fibers (Fig. 2D-F).

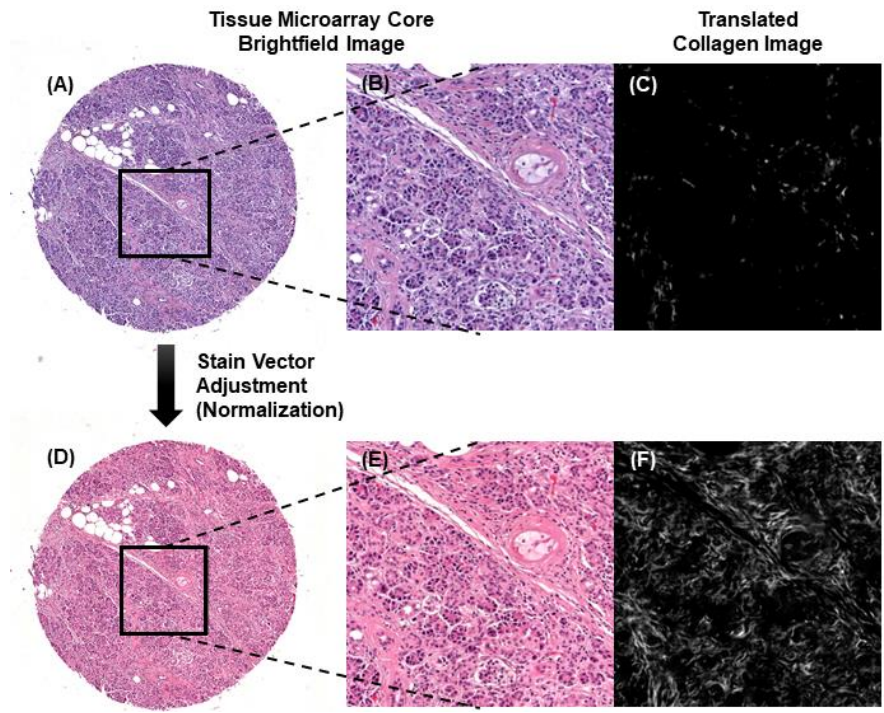

*Figure 2. Normalization of stain vectors. The top row shows a brightfield image of H&E-stained pancreatic TMA core from the dataset before stain normalization (A). The ROI represented by the black box is magnified (B) and corresponding translated collagen image is shown (C). The bottom row shows the same core (D) with the same ROI (E) and corresponding translated collagen image (F) but, after the post stain vector adjustment.*

Figure 3 demonstrates the comparison of translated collagen images with SHG images of the same stroma. The upper row displays the H&E-stained BF image (Fig. 3A), the generated collagen image (Fig. 3B), and the SHG-based collagen image (Fig. 3C). SSIM index indicated good similarity between the translated-real collagen image pair (0.578). The fiber orientations and alignment of translated-real collagen image pair was calculated by the curvelet method using the CurveAlign software. Boxplots of fiber orientation values reveal the statistical similarity (Fig. 3D). The heatmap of fiber orientations of the SHG collagen image (Fig. 3E) and an equivalent heatmap of the generated collagen image (Fig. 3F) were observed. In the heatmap image, red indicates an orientation of 60 degrees or more, yellow shows 45 to 60 degrees, green denotes 10 to 45 degrees, and grey represents angles of 0 to 10 degrees. A Bland-Altman plot of the sine values of fiber orientation shows a very good agreement between translated and SHG collagen image (Fig. 3G). The Pearson correlation coefficient (0.95) computed between the values of alignment showed excellent agreement (Fig. 3H). It was verified using the Bland-Altman graph (Fig. 3I).

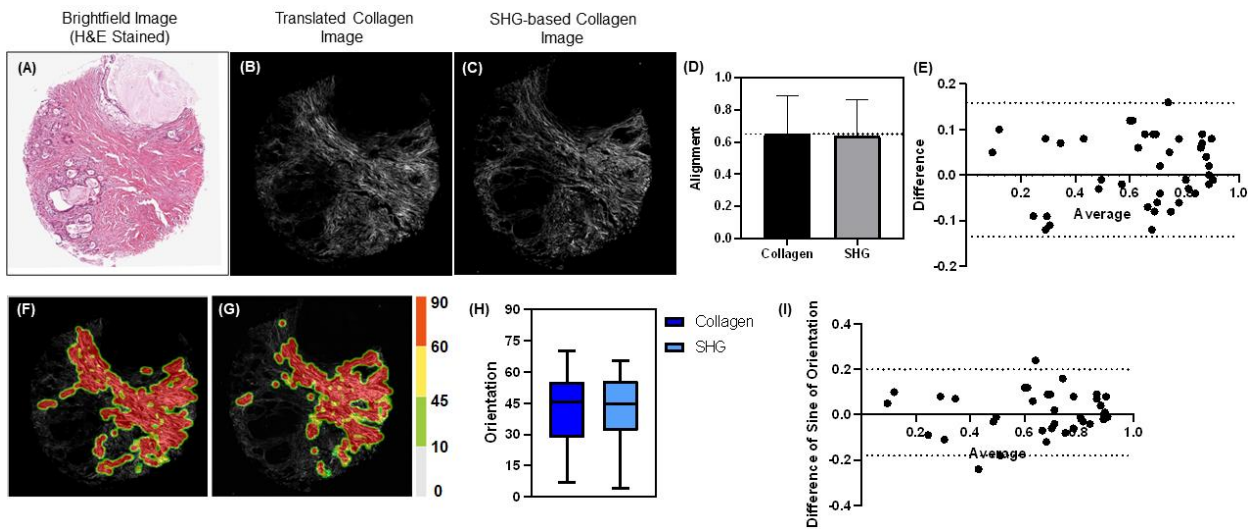

*Figure 3. Translated collagen image is equivalent to real collagen image. The brightfield image of H&E-stained pancreatic stroma (A) is computationally translated to the collagen image (B), which is equivalent to the second harmonic generation (SHG) based real collagen image of the same stroma (C). The alignment values extracted using CurveAlign showed no significant difference between translated and real images (D). The similar result was shown by their Bland-Altman plot (E). The orientation heatmaps (F, G) computed by CurveAlign are overlaid on the translated collagen image (B) and the real collagen image (C). The boxplot of fiber orientations (H) of translated (blue) and real (cyan) collagen fibers shown no significant difference. The Bland-Altman plot (I) of sine values of orientations corroborates the result shown (H).*

## b.    Alteration of Collagen Morphology and Organization in Pancreatic Diseased Conditions

Using the neural network, we translated the stain-normalized BF images of pathology-reviewed clinical tissue cores to collagen images, resulting in submicron-resolution data about collagen organization and topology. Figure 4 displays the brightfield image of stain normalized tissue core, the corresponding computationally generated collagen image, as well as the extracted collagen fibers and orientation map. In these images, the normal pancreatic cores (Fig. 4A) and the cores of CP (Fig. 4B) show a lesser extent of fibrosis with reduced alignment in comparison with the cores of PanIN (Fig. 4C) and PDAC (Fig. 4D)

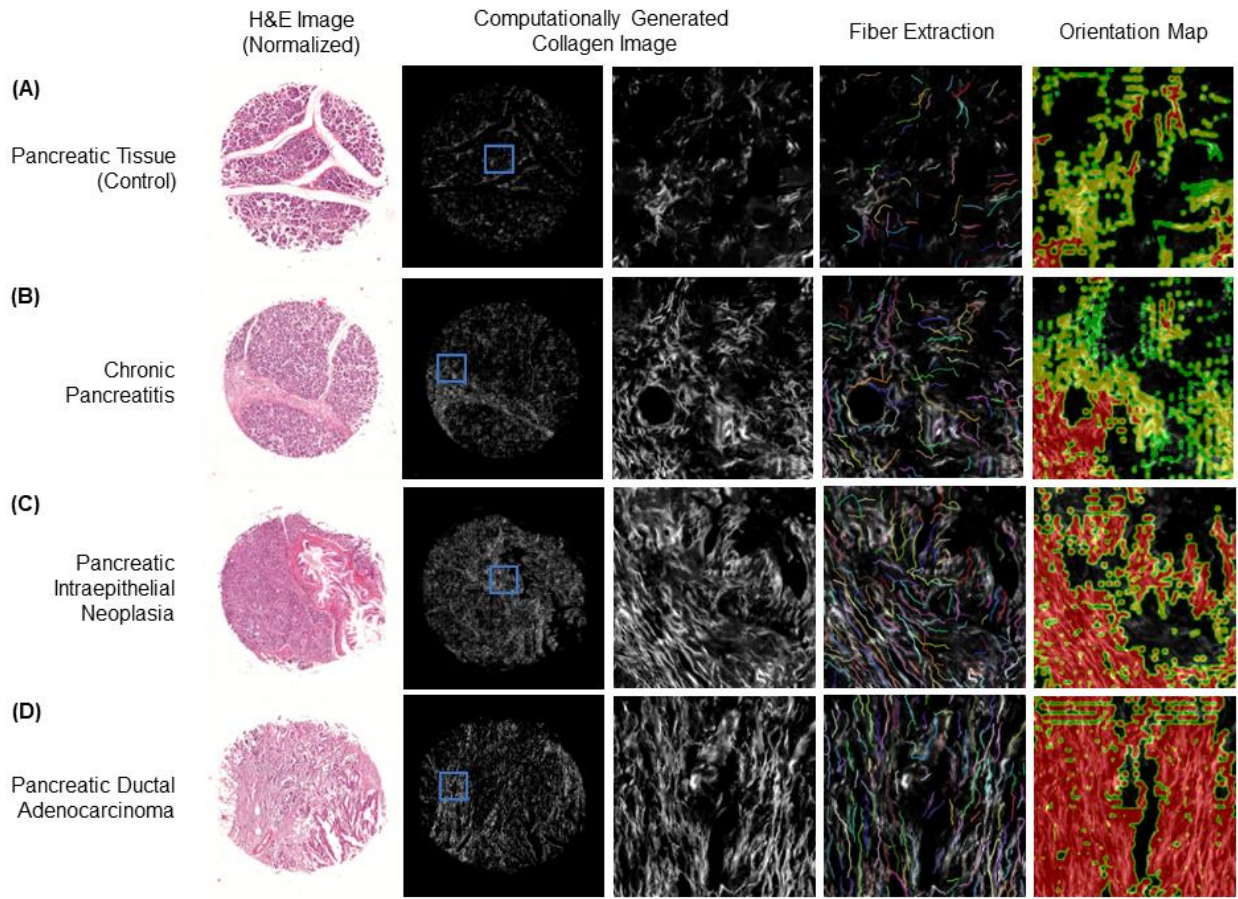

*Figure 4. Visualization of collagen remodeling in translated collagen images. The stain normalized brightfield images of four classes of pancreatic TMA cores analyzed in the study include normal pancreatic tissue (A), chronic pancreatitis (B), intraepithelial neoplasia (C) and ductal adenocarcinoma (D); an ROI from each computationally generated collagen images (indicated by blue box) show the extent of fibrosis. The collagen fibers were extracted using CT-FIRE and were overlaid on each translated image. Similarly, the orientation heatmap computed using CurveAlign were also overlaid on each translated image.*

We calculated and evaluated the mean values of the collagen parameters extracted from the four categories of tissue cores. The plots shown in Figure 5 compare the collagen characteristics revealing significant differences between collagen fibers of the four groups. There was a significant difference in fiber length among all four groups, with fibers in PDAC and PanIN cores being more elongated than normal pancreatic tissue and CP (Fig. 5A). There was also a statistically significant increase in the width of fibers in PanIN cores as compared to normal pancreatic tissue and CP (Fig. 5B). The results for density calculation showed that fibers in normal tissue cores were significantly less closely packed than the other three groups (Fig 5C). The alignment results showed that fibers in PDAC cores were significantly more aligned than all other three groups (Fig. 5D). The collagen deposition in normal cores was also significantly lesser than in PanIN and PDAC cores (Fig 5E). The k-dispersion parameter results showed that fibers in PDAC cores were significantly higher values than all other three groups (Fig. 5F). The results of statistical ANOVA for all involved collagen parameters, including the cumulative F-value, P-value and R-square, are shown in Table 1.

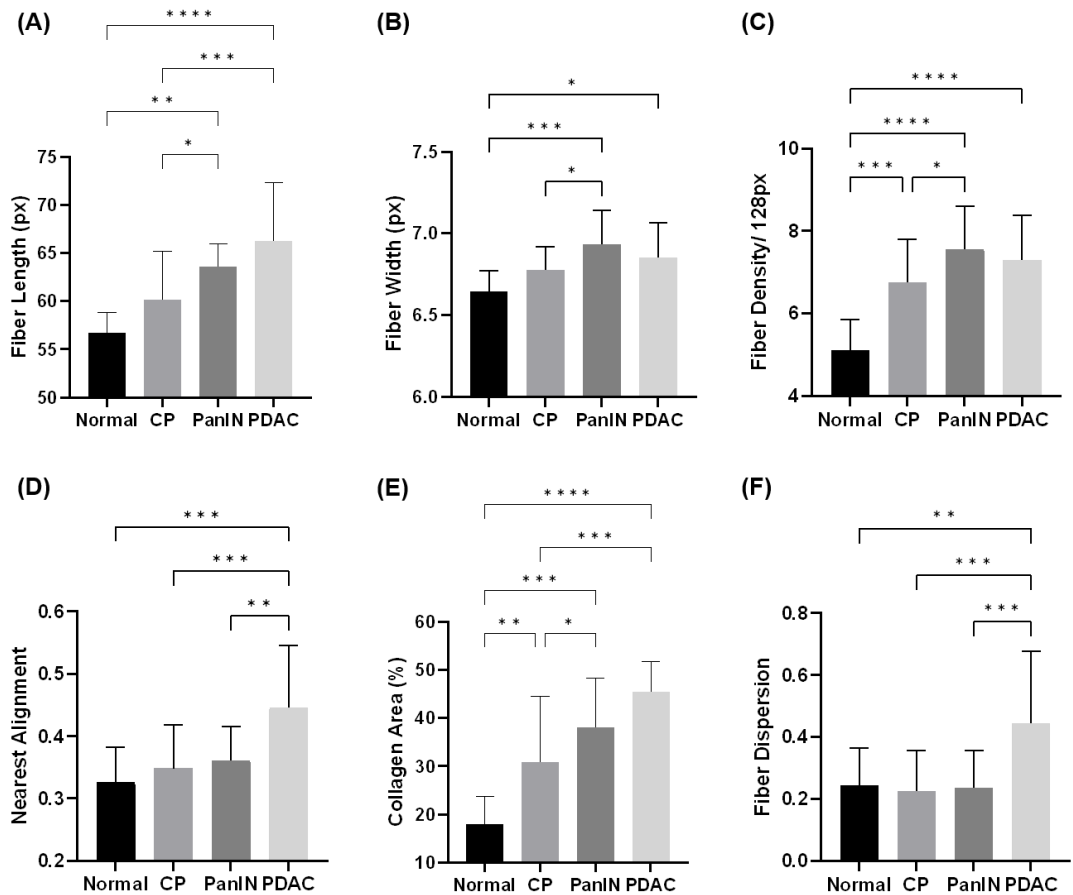

*Figure 5. Quantitative collagen remodeling in pancreatic lesions. CT-FIRE and CurveAlign were used to quantify structural and organizational parameters of collagen fiber i.e., length (A), width (B), density (C) and, alignment (D). The collagen area (E) was computed using Otsu's thresholding while the dispersion factor k (F) was calculated using FiberFit tool. Results are shown as mean ± SD for the four classes of pancreatic TMA cores. N = 60 [8 Normal, 24 CP, 15 PanIN, 13 PDAC]. (\*p<0.05, \*\*p<0.01, \*\*\*p<0.001, \*\*\*\*p<0.0001.).*

*Table 1. Comparison of collagen morphological parameters of the four pancreatic tissue classes.*

| Collagen Parameters | Control | CP | PanIN | PDAC | F value | P value | $R^2$ |
|---|---|---|---|---|---|---|---|
| Fiber Length | 56.76 | 60.13 | 63.57 | 66.33 | 9.066 | <0.0001 | 0.335 |
| Fiber Width | 6.643 | 6.778 | 6.933 | 6.854 | 5.06 | 0.0037 | 0.2194 |
| Fiber Box Density | 5.103 | 6.747 | 7.559 | 7.298 | 9.988 | <0.0001 | 0.3569 |
| Nearest Alignment | 0.3257 | 0.3483 | 0.3613 | 0.4462 | 6.304 | 0.001 | 0.2594 |
| Collagen Area | 21.83 | 30.8 | 38.09 | 45.48 | 7.665 | 0.0002 | 0.2987 |
| Dispersion Parameter | 0.1529 | 0.1083 | 0.1413 | 0.2569 | 7.417 | 0.0003 | 0.2918 |

Based on the feature space formed by each of the extracted collagen parameters, we implemented the KNN classifier to discriminate the neoplastic and non-neoplastic tissues. The PanIN and PDAC tissue cores were grouped as neoplastic, while the CP and normal pancreatic tissue cores were categorized as non-neoplastic. The plots displayed in Figure 6 and their corresponding values in Table 2, display the performance of each parameter in distinguishing the cores. Among the six collagen parameters, the area of collagen deposition and the fiber length showed higher accuracy with values of 0.793 and 0.776, respectively. The same parameters also had larger AUCROC values of 0.83 and 0.81, respectively, indicating their effectiveness in identifying a distinct collagen topology in the neoplastic stromal microenvironment.

*Table 2. Receiver operating characteristics of collagen morphological parameters distinguishing neoplastic (PanIN and PDAC) and non-neoplastic (Normal and CP) tissues.*

| Collagen Parameters | AUC | Youden's Index | Sensitivity | Specificity | Accuracy |
|---|---|---|---|---|---|
| Fiber Length | 0.81 | > 65 | 0.8929 | 0.6667 | 0.776 |
| Fiber Width | 0.78 | > 6.8 | 0.5357 | 0.9 | 0.724 |
| Fiber Box Density | 0.74 | > 0.41 | 0.7143 | 0.7333 | 0.693 |
| Nearest Alignment | 0.77 | > 6.35 | 0.6429 | 0.8 | 0.724 |
| Collagen Area | 0.83 | > 40 | 0.8214 | 0.7669 | 0.793 |
| Dispersion Parameter | 0.77 | > 0.19 | 0.8571 | 0.5667 | 0.7068 |

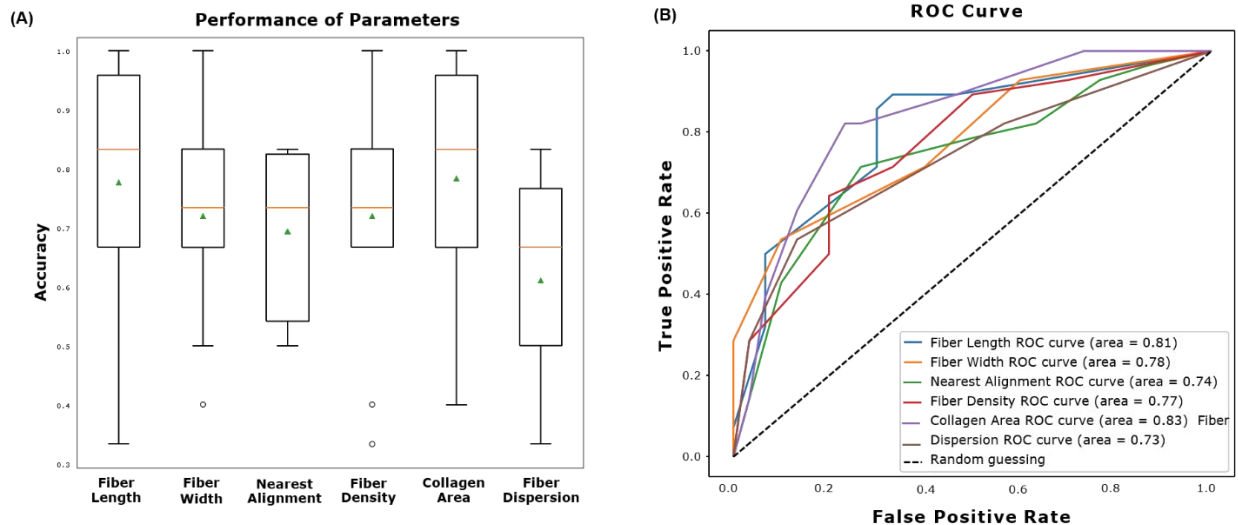

*Figure 6. Performance of collagen morphological parameters in classification of neoplastic (PanIN and PDAC) and non-neoplastic (Normal and CP) tissues. The image on the left shows average testing accuracies for each parameter. The image on the right displays the receiver operating characteristic curve of each parameter and their respective area under the curve.*

## Discussion

The integration of stromal biomarkers for diagnostic purposes is has not been receiving adequate attention from pathologists, and their significance remains not well acknowledged. In addition, SHG microscopy, which is utilized to analyze and quantify stromal collagen fibers, is not feasible for use in routine practice in the current scenario because it is expensive and needs specialized equipment and technical knowledge that are not easily accessible in standard diagnostic laboratories. However, clinical practice has recently begun to identify the stroma as a crucial component in tumor pathogenesis. Collagen associated features such as orientation around the tumor, quantity of deposition, changes in relative alignment etc. have been demonstrated to be predictive indicators in several investigations[19-28]. It is expected that in the near future the relevant research works especially clinical prospective studies might be benefited by better assessment of altered morphological and organizational characteristics of fibrillar collagen, if directly implemented in the workflow of routine histopathology.

A critical challenge in histopathology is the color variation of the stained slide across various scanners and laboratories. Stain normalization has the capacity to lower stain variability and enhance the reliability of image assessment techniques and computer aided diagnosis. As observed in Figure 2, normalization of the stain vectors improved the image translation process and aided in better visualization of the collagen fibers. The use of histogram modification enabled the normalization of stain vectors in these images. However, for a larger image database having more diverse stain vectors, a more comprehensive approach might be required.

The collagen image translated using CNN are capable of detecting the changes in architecture of the collagen network. The divergence from perfect correlation with real SHG images, however, is influenced by a variety of contributing factors. One of these factors is SHG shot noise which can compromise single collagen fibers due to a lack of discernible SHG signal, whereas translated images display a smoother fiber ridge, readily identifiable as an unbroken fiber. The other factor is that the resolution of collagen fibers in the translated images is highly dependent on the numerical aperture (NA) of the microscope objective. Additionally, the high photon density requirement of SHG imaging may affect visualization of extremely tiny collagen fibers, or, the fibers existing in close proximity might be depicted as a single fiber. Also, antiparallel alignment of fibers may weaken or even cancel out the SHG signal, which potentially can lead to less accurate pixel-wise fiber extraction. However, these concerns do not occur in BF imaging as it involves first order optical phenomena. The execution of image translation goes beyond classification based on color or texture by training the CNN to accurately anticipate collagen fibers in the tissue. This training enables the translated image to have a good agreement and similarity (Figure 3) with real SHG image making it a valuable tool for researchers in a wide array of collagen remodeling studies.

The impact of collagen remodeling on the onset and progression of PDAC is widely recognized[70]. In order to investigate this phenomenon, fiber extraction and quantitative characterization of collagen remodeling were performed on translated whole core images. Significant increase in the length, density, and area of deposition of fibrillar collagen were observed in PanIN and PDAC tissues compared to the normal tissue. However, there were no significant differences observed in these parameters between PanIN and PDAC. Combining both results together implies that collagen associated features in the precursor lesion (i.e., PanIN) may serve as an indicator of the onset of tumor progression[71]. Further, two different approaches were applied to analyze the alignment of the collagen network: the curvelet transform and the Fourier transform. The results of both approaches were observed in agreement with each other and showed a significant difference in alignment between PDAC and the other three classes of pancreatic tissues, indicating an increase in the anisotropy of the collagen network in PDAC. An enhancement in relative alignment of stromal collagen emphasizes the significance of fibrillar collagen as a stromal biomarker for progression of PDAC from PanIN, as observed in multiple earlier studies[72,73].

The role of pancreatic stellate cells (PSCs) in alteration of stromal collagen was reported in numerous studies[74,75]. Low-grade PanINs exhibit a significant increase in periductal collagen deposition, despite the absence of a large increase in PSC activation. As PanINs advances to higher-grade lesions, the density of collagen remains constant, but there is a noticeable increase in the activation of PSCs. Studies have highlighted the critical role of PSCs in the progression of PDAC and their function as the primary source of fibrillar collagens[76]. In-vitro studies have revealed that highly contractile PSCs create aligned collagen networks, which can promote cancer cell migration in PDAC[77]. This implies that the augmented activation of PSCs could be responsible for the collagen remodeling especially increase in alignment as observed in PDAC. Conversely, the reduced activation of PSCs could potentially explain the absence of a significant increase in collagen fiber alignment in low-grade PanIN and CP[78].

Studies have shown the impact of collagen fiber alignment on cancer cell by altering their behavior and ability to invade the surrounding tissues. Alignment can significantly impact the way genes are expressed, and the way cancer cells differentiate, proliferate, and metastasize. Collagen fibers with increased alignment act as "highways" for cancer cell movement. Moreover, the remodeling of collagen fibers can affect the porosity of the ECM, potentially restricting cell movement within the tumor and enhancing the ability of cancer cells to invade surrounding tissues for their survival. The altered alignment of collagen fibers was reported as a key factor in the metastatic invasion of multiple carcinomas and might be useful in determining local invasion capacity[79,80]. Therefore, an early and accurate assessment of pancreatic stromal collagen can aid in an enhanced understanding of different stages of carcinogenesis, ultimately could assist in an improved patient prognosis.

The whole core images utilized in the study can provide a more comprehensive view of the tissue and has more clinical relevancy. Our study demonstrated a quantitative collagen assessment of pancreatic lesions in the translated whole core images rather than selection and annotation of specific ROIs in the given image. The use of whole core images can facilitate an automated approach for disease detection, potentially alleviating the reliance on clinical expertise. Additionally, taking the full extent of the tissue into consideration may increase the generalizability of the algorithm to analyze collagen remodeling by reducing the potential for sampling errors and selection bias. Overall, such approach has high potential to improve the efficiency of pathologist and even enhance the predictive accuracy if implemented in histopathological workflow analysis[81].

While in recent several reports have been demonstrated the quantitative analysis of fibrillar collagen in pancreatic tissues but, majority of them are using the method of SHG microscopy. In this study, we have demonstrated that quantitative analysis of collagen in the pancreatic stroma can be performed using the cross-modality translation of H&E-stained BF images. Further, extracted collagen fibers from the translated images were quantitatively characterized and a difference between the stromal collagen characteristics in pancreatic lesions were observed. Our results indicate that the amalgamation of CNN-based cross-modality image translation and quantitative image analysis approaches – comprising a customized combination of FiberFit, CT-FIRE, and CurveAlign – in a brightfield microscope could potentially enable the automated screening of collagen remodeling in pancreatic lesions. A quantitative and accurate assessment of collagen architecture can further complement the conventional histopathological insights, changes in tissue microenvironments and, motivate to further explore the possibility to use fibrillar collagen as a potential stromal biomarker for pancreatic lesions.

## Acknowledgements

We gratefully acknowledge the funding from Department of Biotechnology (DBT), Government of India. This work was supported by Ramalingaswami Re-entry Fellowship.